# Suspended graphene: a bridge to the Dirac point.

Xu Du, Ivan Skachko, Anthony Barker, Eva Y. Andrei

Department of Physics & Astronomy, Rutgers the State University of New Jersey

**The recent discovery of methods to isolate graphene[1-3], a one-atom-thick layer of crystalline carbon, has raised the possibility of a new class of nano-electronics devices based on the extraordinary electrical transport and unusual physical properties[4,5] of this material. However, the experimental realization of devices displaying these properties was, until now, hampered by the influence of the ambient environment, primarily the substrate. Here we report on the fabrication of Suspended Graphene (SG) devices and on studies of their electrical transport properties. In these devices, environmental disturbances were minimized allowing unprecedented access to the intrinsic properties of graphene close to the Dirac Point (DP) where the energy dispersion of the carriers and their density-of-states vanish linearly giving rise to a range of exotic physical properties. We show that charge inhomogeneity is reduced by almost one order of magnitude compared to that in Non-Suspended Graphene (NSG) devices. Moreover, near the DP, the mobility exceeds 100,000 cm$^2$/Vs, approaching theoretical predictions for evanescent transport in the ballistic model.**

The low energy excitation spectrum of graphene mimics relativistic particles - massless Dirac fermions (DF) - with an electron-hole symmetric conical energy dispersion and . vanishing density of states at the DP. Such unusual spectrum is expected to produce novel electronic properties such as negative index of refraction[6], specular Andreev reflections at graphene-superconductor junctions[7,8], evanescent transport[9], anomalous phonon softening[10], etc. A basic assumption behind these intriguing predictions is that the graphene layer is minimally affected by interactions with the environment. However in reality the environment[11,12] and in particular the substrate[13], can be quite invasive for such ultra-thin films. For example, the carrier mobility in graphene deposited on a substrate such as $Si/SiO_2$ deteriorates due to trapped charges in the oxide or to contaminants that get trapped at the graphene-substrate interface during fabrication. The substrate-induced charge inhomogeneity is particularly deleterious near the DP where screening is weak,[14,15] leading to reduced carrier mobility there. In addition, the atomic roughness of the substrate introduces short range scattering centers and may contribute to quench-condensation of ripples within the graphene layer[16].

In order to eliminate substrate induced perturbations, graphene films were suspended from Au/Ti contacts to bridge over a trench in a $SiO_2$ substrate. In contrast to prior realizations of suspended graphene[17,18] which did not provide electrical contacts for transport measurements, the SG devices described here incorporate multiple electrodes that allow 4-lead transport measurements. The SG devices employed here were fabricated from conventional NSG devices using wet chemical etching (see supporting online material). In a typical SG device, shown in Figure 1b, the graphene layer is suspended from the voltage leads which at the same time provide strong structural support. Such structure avoids complications, such as the ultra-sensitivity to details of lead geometry, that arise when transport measurements are carried out with a conventional Hall-bar configuration in ballistic devices[19]. In contrast, for the two-lead voltage configuration used here, the measured transport properties of ballistic devices depend on lead separation and doping in a straightforward way that can be calculated from first principles[8].

Low temperature (4.2K) magnetotransport measurements were carried out to determine the number of graphene layers and to obtain the relation, $n(V_g)$, between the induced carrier density, $n$, and the gate voltage $V_g$. From the gate dependence of the Shubnikov-deHaas (ShdH) oscillations (Figure 1c) a "fan diagram" (Figure 1d) was constructed for the hole branch, by plotting the Landau Level index vs. $1/B$ for several values of $V_g$. The common intercept of all the curves at $y = \frac{1}{2}$, signaling a Berry phase of $\pi$, proves that the sample is a single layer of graphene[20]. The slopes of these curves give the ShdH oscillation periods, $1/B_F$, for each value of $V_g$ from which we then obtain the gate dependence of the carrier density $n = \frac{4e}{h} B_F(V_g)$, shown in Figure 1e. We note that $n(V_g)$ is linear in $V_g$ indicating that, within the range of applied gate voltages, bending of the graphene sheet due to the electrostatic force is negligible. Comparing the "carrier density capacitance" of the SG sample $\alpha_{SG} = n/V_g$ (2.14x10$^{10}$cm$^{-2}$/V) to that of the NSG sample, ($\alpha_{NSG} \sim 7.4 \times 10^{10}$ cm$^{-2}$/V) we find that the ratio $\alpha_{NSG}/\alpha_{SG} = 3.46$ is ~15% smaller than the ratio of the corresponding dielectric constants (SiO$_2$ and vacuum). This suggests a slight "permanent sagging" of the SG devices attributed to the deformation of the leads by the wicking action of liquids during the fabrication process.

We next focus on the carrier density dependence of the resistivity, $\rho(n)$, in zero magnetic field in the temperature range 4.2K to 250K (Figure 2a). We note that the resistivity peak almost coincides with the neutrality point ($V_g = 0$) indicating little extrinsic doping. The asymmetry in the curves, which was observed in all our SG samples can be attributed to the invasiveness of the metallic contacts which primarily affect the electron branch[21] (see online supporting material). We will limit our discussion to the hole branch where well-defined ShdH oscillations are observed. For the SG samples we note that the peak in $\rho(V_g)$ at the DP becomes very sharp at low temperatures, and on the hole branch the half width at half maximum (HWHM) at the lowest temperature ($\delta V_g \sim 0.15$ V, $\delta n \sim 3.2 \; 10^9 cm^{-2}$ for the sample with channel length $L = 0.5$ µm) is almost one order of magnitude narrower than that of the best NSG samples published so far[4,22]. This is directly seen (Figure 2a) in the side by side comparison of the $\rho(n)$ curves for SG and NSG samples of the same size and taken from the same graphite crystal. The HWHM of the NSG sample shown here is $\delta V_g \sim 3V$ ($\delta n \sim 2.2 \; 10^{11} cm^{-2}$). In Figure 2b we compare the low temperature $\rho(n)$ curves with ballistic model predictions for two SG samples. Although the ballistic curve (HWHM $\delta V_g \sim 0.01V$, for $L = 0.5$µm) is still sharper than the SG curve, the discrepancy is much smaller than for NSG samples. Furthermore, we note that the maximum resistivity values SG devices (12KΩ for the sample shown) are significantly higher than the typically observed 3~7 KΩ values in NSG samples. At low carrier densities, the suspended graphene devices show near-ballistic transport behavior away from the DP, as revealed by the weak graphene channel length dependence of the conductance per unit width shown in Figure 2c.

A remarkable feature of the $\rho(n)$ curves in SG samples is the strong temperature dependence of the maximum resistivity at the DP, in stark contrast to NSG samples. Whereas in NSG samples the maximum resistivity saturates below $\sim 200K^{23}$, in SG samples it continues to increase down to much lower temperatures. We attribute this difference to reduced charge inhomogeneity in the SG samples. The charge inhomogeneity introduces electron-hole puddles[14] close to the DP which cause spatial fluctuations in doping levels. These fluctuations define an energy bandwidth, $E_f^{sat}$, for the average deviation of local DP from the Fermi energy, $E_f$. Within this energy range: $|E_f| < E_f^{sat}$, the effect of gating is limited to a redistribution of carriers between electrons and holes without significantly changing the total carrier density. Similarly the effect of thermally excited electron-hole pairs is masked until $k_B T \sim E_f^{sat}$ ($k_B$ is the Boltzmann constant). The effect of charge inhomogeneity can be estimated from the carrier density dependence of the conductivity, $\sigma(n) = \rho^{-1}(n)$, near the DP by finding the saturation carrier density, $n^{sat}$, below which the conductivity tends to a constant as illustrated in Figure 3a. In Figure 3b we compare the temperature dependence of the corresponding energy scale, $E_F^{sat} = \hbar v_F \sqrt{n^{sat} \pi}$ for two SG samples and one NSG sample. ($v_F = 10^6$ m/s is the Fermi velocity). At high T, the slopes of both SG curves approach $k_B$ as expected for thermally excited carriers. At the lowest temperatures where the fluctuation energy is controlled by charge inhomogeneity, its value in the SG samples, $E_F^{sat} < 10 meV$, is much smaller than in the best conventional NSG samples $\sim 40 meV$ [22]. The effect of temperature on the transport properties of SG is best illustrated by considering, $\sigma(n)$, the carrier density dependence of the conductivity. We find that outside the saturation regime and for T<100K, the data depend only weakly on temperature and coincide with the theoretical curve when the latter is scaled down by a factor of 2.2. This observation indicates that for T<100K the scattering is roughly independent of *T*. By contrast the minimum conductivity, $\sigma_0$, seen at the DP shows strong T dependence. This can be quantified by plotting $\sigma_0(T)$ (Figure 3c inset). The linear *T* dependence of $\sigma_0(T)$ seen for *T*<100K can be understood in terms of thermally excited carriers and scattering that is independent of T. The finite intercept at *T*=0 can be attributed to a combination of evanescent transport and e-h puddles. For T>100K, the sub-linear T-dependence of $\sigma_0$ suggests an onset of a thermally excited scattering mechanism. Away from the DP and for T>100K, strong T-dependence is observed throughout the whole carrier density range, suggesting thermally a activated scattering mechanism.

A direct consequence of the low level of charge inhomogeneity in the SG samples is that one can follow the intrinsic transport properties of Dirac fermions much closer to the DP than is possible with any NSG samples fabricated to date. In Figure 4a we compare the carrier density dependence of $\sigma/e\alpha V_g$ for SG and NSG samples. Outside the puddle regime, $\alpha V_g > n^{sat}$ ($n^{sat}$ marked by arrows) this measures the carrier mobility $\mu = \sigma/e\alpha V_g = \sigma/en$. For $n < n^{sat}$ however, this expression cannot be interpreted as mobility. At low carrier densities (just outside the puddle regime) we find that the maximum mobility of the SG samples exceeds 100,000 cm$^2$/Vs. compared to ~ 2,000-20,000 cm$^2$/Vs in the best NSG samples. Since at low densities the mobility is mostly determined by long range Coulomb scattering[16] (short range scattering is very weak near the DP due to the small density of states[24]), the difference in mobility between the SG and NSG samples is naturally attributed to substrate-induced charge inhomogeneity. In this picture the removal of the substrate in the SG samples eliminates the primary source of Coulomb scattering, the trapped charges. At high carrier densities ($n > 4 \times 10^{11}$ cm$^{-2}$), the mobility in the two types of samples becomes comparable (~10000 cm$^2$/Vs) indicating that short range scattering becomes dominant there. The short range scattering can be attributed to imperfections in the graphene layer which may reflect defects in the parent graphite crystal or could be introduced during the fabrication process. Both sources of defects can in principle be reduced to produce SG samples with even better quality.

To further shed light on the dominant scattering in the two types of samples we compare in Figure 4b the carrier density dependence of the mean free path: $mfp = \dfrac{\sigma h}{2e^2 k_F}$. In the NSG sample the mfp increases with carrier density for $n^{sat} < n < 4 \times 10^{11}$ cm$^{-2}$ reflecting dominant Coulomb scattering[16] there. This trend which is not seen in SG samples, again suggests almost complete absence of Coulomb scattering[16,24].

In Figure 4d we focus on the temperature dependence of the mfp($n$) curves for the SG sample. The negative slope and absence of T dependence for T < 100K, suggest that scattering is predominantly from short range scatterers. For T>100K however, the slopes of the mfp($n$) curves become increasingly positive, suggesting thermally induced long range scattering. Such long range scattering cannot be attributed to Coulomb scattering from charged impurities because such mechanism is expected to be independent[23] of T. One possibility is that for $T$>100K thermally excited graphene ripples[16] introduce additional scattering, However, more work is needed to fully understand the scattering mechanism in this regime.

Finally, we compare transport at the lowest temperatures, where thermally activated scattering is absent, with the theoretical prediction for a ballistic junction[8]. According to the ballistic model[8] the ρ(n) curve in graphene devices is very sharp near the DP. Although, as seen in Figure 2b, our SG samples do not precisely follow the theoretical predictions, they approach the ballistic limit when gated close the DP. There the values of the mfp in the SG devices are within ~50% of the ballistic model prediction, indicating that there are still some scattering events in the graphene channels.

In ideal ballistic graphene junctions, the transport at the DP is dominated by transmission of evanescent modes[9]. Previously measured values for the maximum resistivity near the DP in NSG samples (mostly between 2 ~ 7 KΩ) were significantly lower than the theoretical prediction: $\frac{\pi h}{4e^2} \sim 20 K\Omega$ [4,8]. We have shown here that the maximum resistivity for SG samples (12 ~ 40 KΩ) approaches the value expected for transport by evanescent modes. However this fact does not, in itself, constitute a proof of evanescent transport. This is because when charge inhomogeneity and short range scatterers are present their contributions to the maximum resistivity have opposite signs: the former diminishes the resistivity while the latter increases it. Therefore, in order to recognize evanescent transport it is also necessary to estimate the effect of these scatterers by comparing the width of the measured ρ(n) peak, which is broadened by all types of scattering, to the calculated width in the ballistic limit. Here we have shown that the ρ(n) peak in SG devices is more than an order of magnitude sharper than in NSG devices. The simultaneous observation of the sharp ρ(n) peak with a maximum resistivity value that approaches the evanescent transport limit provides experimental evidence of near-ballistic transport in graphene.

Acknowledgements

We thank G. Li, Z. Chen for discussions; S.W. Cheong and M. Gershenson for use of AFM and e-beam; V. Kiryukhin for HOPG crystals; F. Guinea, A. Balatsky, M. Fogler and D. Abanin for useful discussions. Work supported by DOE DE-FG02-99ER45742; NSF-DMR-0456473 and ICAM.

Figure captions:

Figure 1: **SG sample characterization. a**. a cartoon showing the structure of the SG devices. **b**. SEM image of a typical SG sample. **c.** ShdH oscillations in the hole branch at indicated gate voltages. The plateaus at $R_{xx}$~13KΩ ($h/2e^2$) reflect a contribution from the Hall resistance seen as a result of the non-ideal shape of the sample. **d.** The "fan diagram": index of the ShdH oscillations (hole-branch) plotted against $1/B$ at indicated gate voltages. The linear dependence extrapolated to the index number ½, identifies the sample as single layer graphene. **e**. Gate induced carrier density obtained from ShdH oscillations plotted against the applied gate voltage.

Figure 2: **Carrier density dependence of transport. a**. Resistivity of SG device with channel length L = 0.5µm and width W= 1.4µm at indicated temperatures (left panel). Resistivity of NSG device with channel length L = 0.5µm (right panel). **b**. Comparison of the carrier density dependence of the resistivity for SG device in the left panel of Fig. 2a with predictions of ballistic model at $T$ = 0. **c.** comparison of conductance per unit width between $L$=0.5 µm (width=1.4µm) sample and $L$=0.25 µm (width=0.9µm)

Figure 3: **Potential fluctuations a**. Comparison of carrier density dependence of conductivity for SG and NSG devices. The arrows indicate the carrier density below which the transport properties of the samples are dominated by potential fluctuations. **Inset**. Similar plot comparing two SG devices (0.5µm and 0.25µm) at $T$ = 4.2K. **b**. Temperature dependence of charge inhomogeneity (measured in amplitude of Fermi

energy fluctuation) for two SG samples (0.5µm and 0.25µm) and one NSG sample (0.5µm). The dotted line is $E_F^{sat} = k_B T$. **c.** Carrier density dependence of the conductivity for indictated temperatures. For T<100, outside the saturation regime and $n < 1 \times 10^{11}$ cm$^{-2}$ all the curves collapse onto the "scaled ballistic" prediction (black curve) which was obtained by dividing the calculated ballistic conductivity by 2.2. In this figure the data were shifted slightly align the minima with the neutrality point. **Inset:** Temperature dependence of $\sigma_0$, the minimum conductivity of the $L$=0.5µm SG device. $\sigma_0$ is dominated by potential fluctuations and thermally excited carriers.

Figure 4: **Mobility and mfp. a**. Comparison of the gate dependence of the mobility $\mu = \sigma/e\alpha V_g = \sigma/en$ for NSG, SG (hole-branch) and ballistic model predictions at $T$ = 100K. The dotted segments at $V_g < V_g^{sat}$, ($V_g^{sat}$ is indicated by arrows) where $\sigma/e\alpha V_g \propto 1/V_g$, characterize the "puddle" regime (charge inhomogeneity) ($V_g^{sat}$ ~ 0.1V for SG and ~1V for NSG). In the "puddle" regime the mobility cannot be obtained from these transport measurements as described in the text. The maximum characterizable mobility for the SG and NSG samples shown are ~120,000 cm$^2$/Vs and 2,0000 cm$^2$/Vs, respectively. The limited mobility and its carrier density dependence shown in the theoretical curve is a result of finite sample size determined by the leads distance. **b**. Gating dependence of the mean free path for NSG, SG samples (hole branch) and ballistic model prediction at $T$ = 100K. **c**. Carrier density dependence of mobility (hole-branch) at indicated temperatures. The dotted parts of the curves indicate the "puddle" regime discussed in Figure 3c where the mobility cannot be obtained from these measurements. The solid parts of the lines show the meaningful part of the data. **d**. Carrier density dependence of mean free path (hole branch) at indicated temperatures.

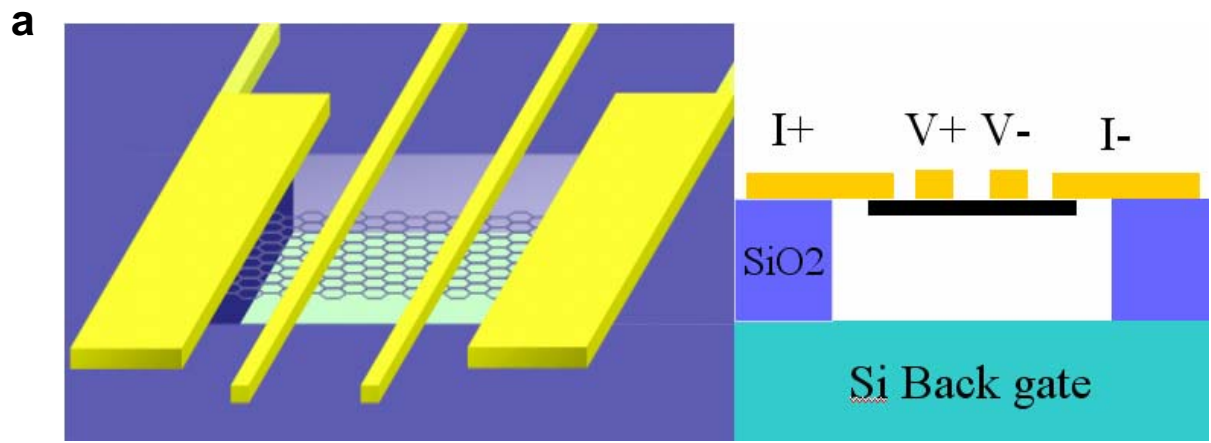

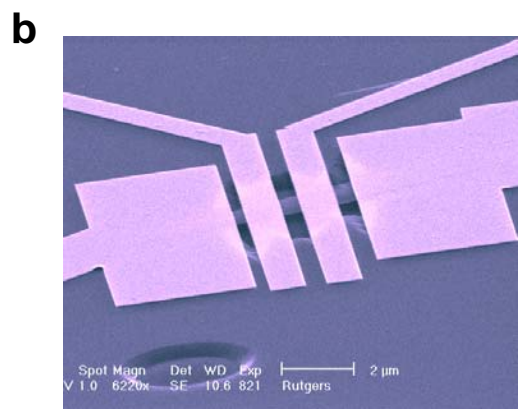
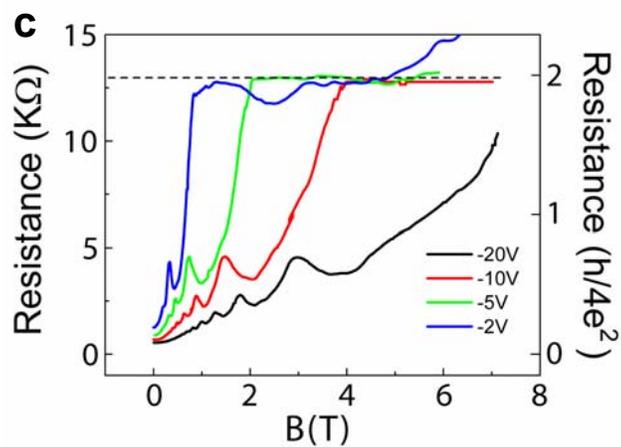

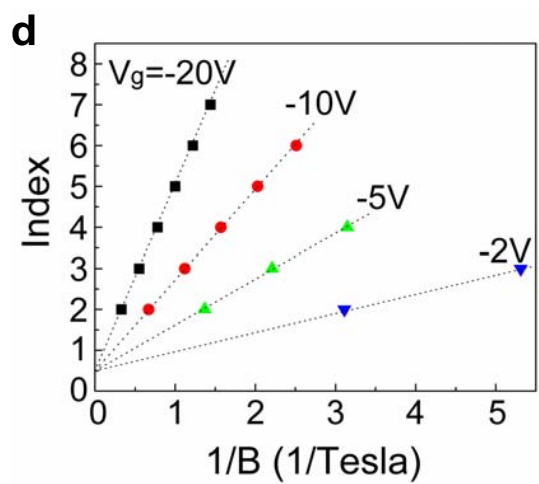
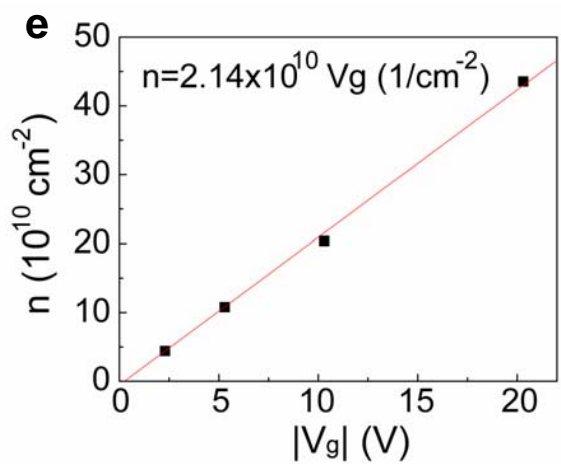

Figure 1

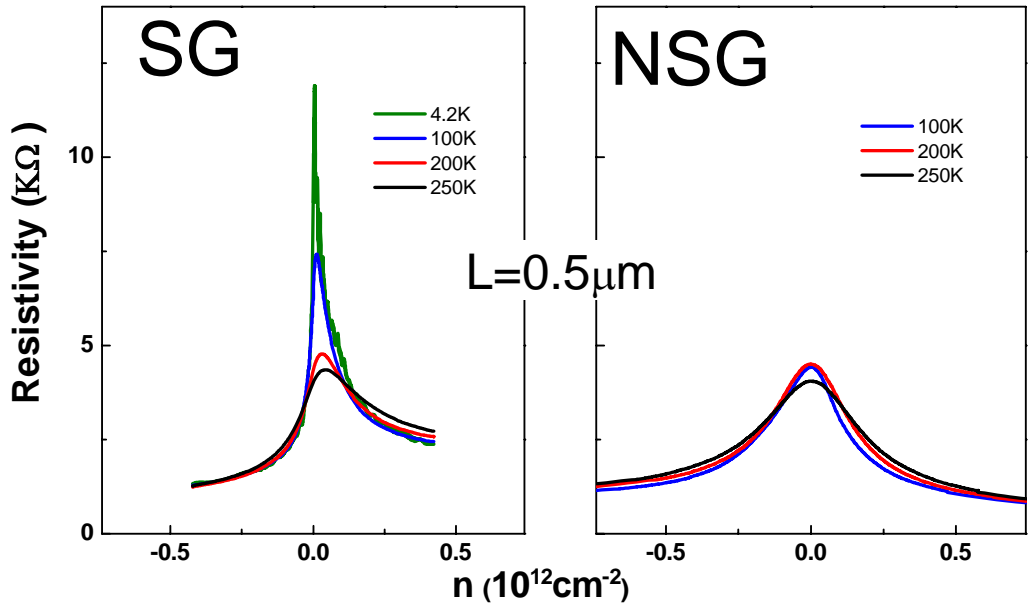
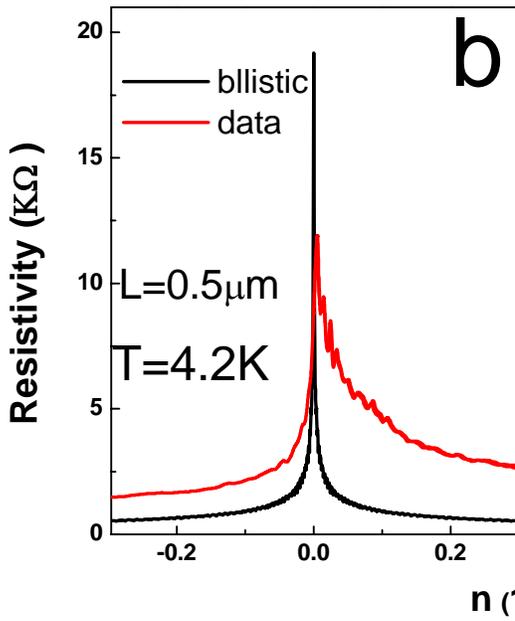
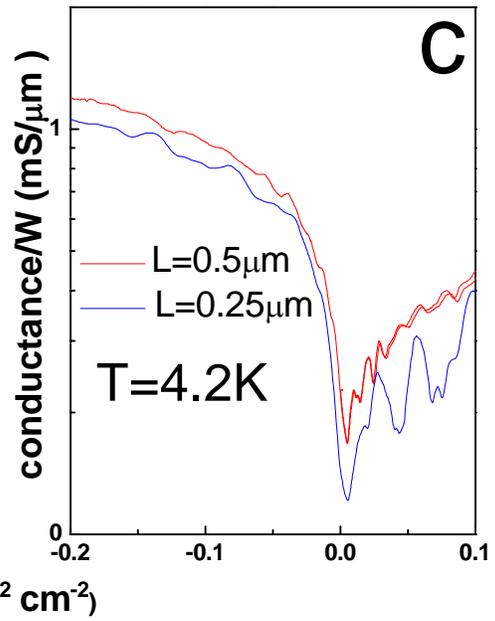

Figure 2

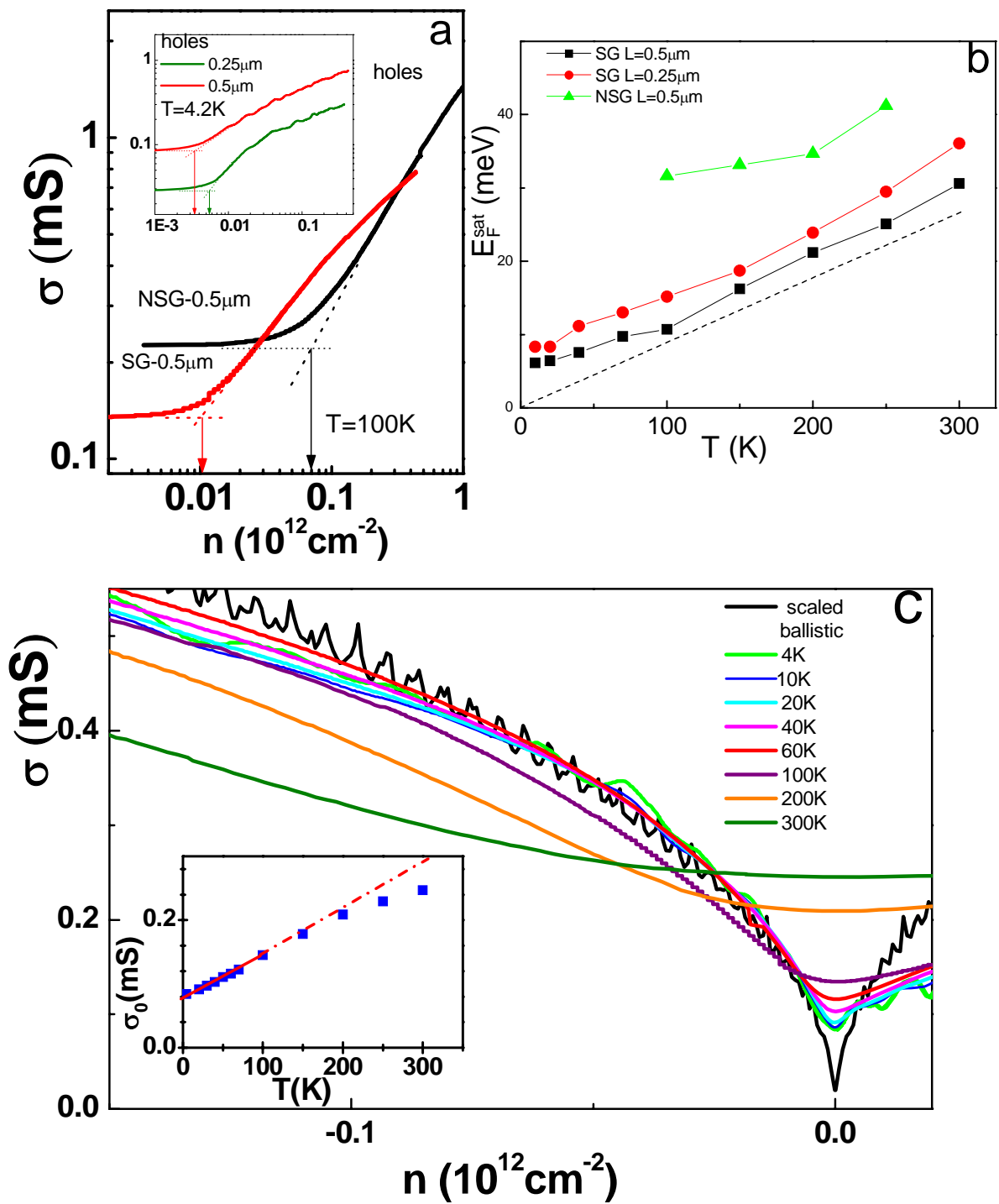

Figure 3

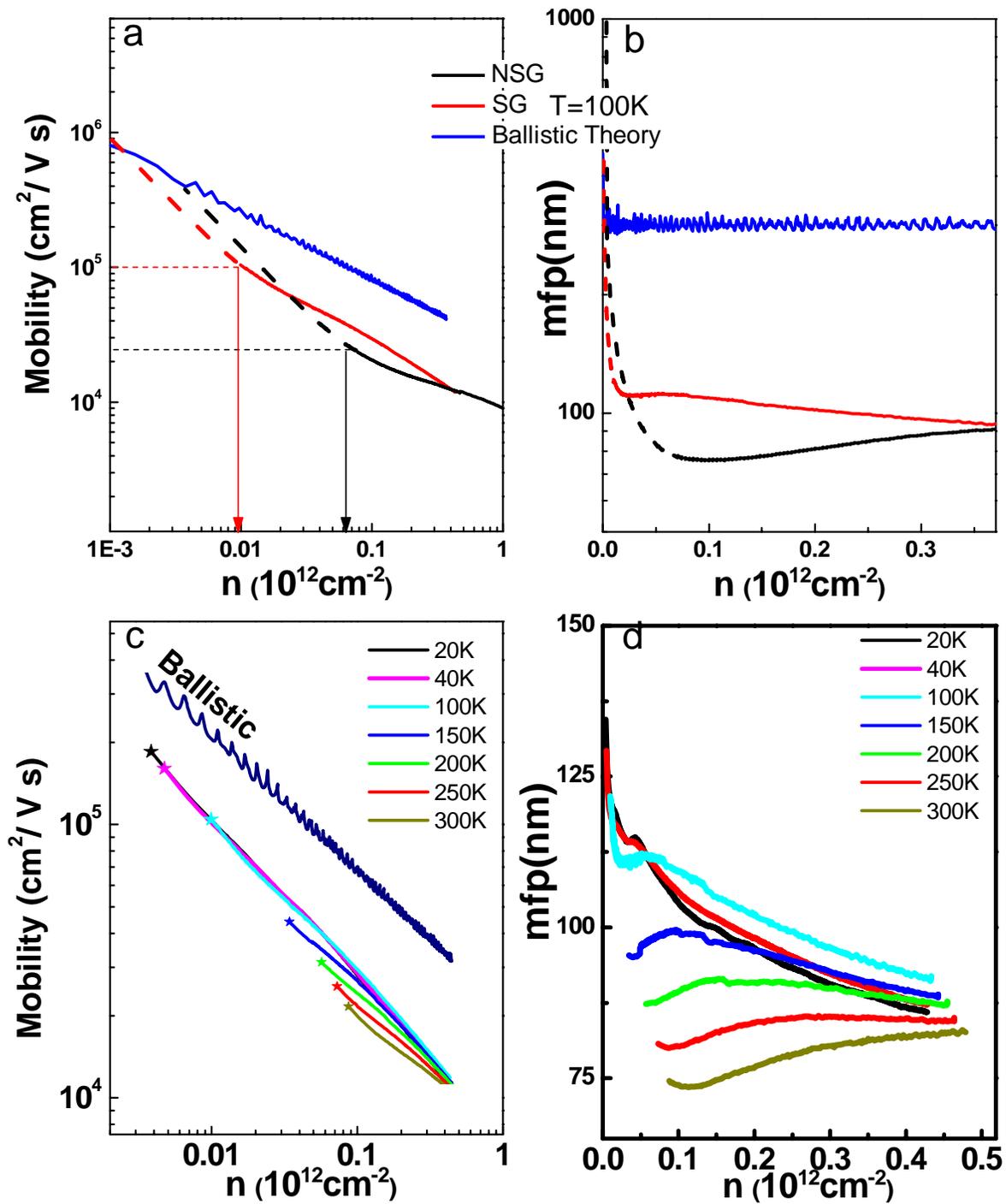

Figure 4